\documentclass[showpacs,prb]{revtex4}
\usepackage{graphicx}
\begin{document}
\title{Synchronization in coupled phase oscillators}
\author{Hidetsugu Sakaguchi}
\affiliation{Department of Applied Science for Electronics and Materials, \\
Interdisciplinary Graduate School of Engineering Sciences, Kyushu
University, 
Kasuga, Fukuoka 816-8580, Japan
}
\begin{abstract}
We make a short review about the synchronization in coupled phase oscillator models. Next, we study the common-noise-induced synchronization among active rotators. At an intermediate noise strength, the noise-induced synchronization takes place most effectively, which is analogous to the stochastic resonance.   Finally, we study the synchronization of coupled phase oscillators with nonvariational interaction on scale-free networks. We find a sharp transition and a weak hysteresis in the nonvariational systems. The sharp transition is found also in the mean-field approximation. 

\end{abstract}
\maketitle
\section{Introduction}
  Limit-cycle oscillation is a typical phenomenon in nonlinear-nonequilibrium systems. 
The synchronization or the frequency locking is a unique phenomenon in a population of coupled limit cycle oscillators.~\cite{rf:1} Huygens observed mutual synchronization of pendulum clocks suspended from the same wooden beam in the 17th century. Buck and Buck reported synchronization of rhythmical flashes of fireflies in South Asia.~\cite{rf:2} Recently, mutual synchronization of cellular activities  was optically observed in the suprachiasmatic nuclei, which is the center of the  circadian rhythm.~\cite{rf:3} Rhythms such as the theta rhythms in the Hippocampus and the gamma rhythms in the visual cortex are considered to play some important roles in information processing in the brain.~\cite{rf:4}     

The theoretical study of the forced entrainment of a limit-cycle oscillator by an external sinusoidal force or the frequency locking between two limit-cycle oscillators has a long history in various research fields such as the classical mechanics and the electrical engineering. In a problem of circadian rhythms, Winfree first studied  mutual entrainment in a population of coupled limit cycle oscillators.~\cite{rf:5}  Kuramoto proposed a soluble coupled phase oscillator model, which has a simple form and exhibits a phase transition analogous to the magnetic phase transition.~\cite{rf:6} 
The Kuramoto model is written as 
\begin{equation}
\frac{d\phi_i}{dt}=\omega_i+\frac{K}{N}\sum_{j=1}^N\sin(\phi_j-\phi_i),
\end{equation}
where $\phi_i$ and $\omega_i$ represent the phase and the natural frequency of the $i$th oscillator, $K$ is the coupling constant and $N$ is the total number of the oscillators. The natural frequency $\omega_i$ is random, and the distribution of $\omega$  is denoted as $g(\omega)$, which is often assumed to be mirror-symmetric with respect to the average value $\omega_0$.  
Kuramoto showed that a phase transition from a disorder state to a mutually synchronized state occurs at a critical coupling strength $K_c$.  He showed that an order parameter $\sigma=(1/N)\sum_j\exp(i\phi_j)$ for the phase order increases continuously from zero as $\sigma\propto (K-K_c)^{1/2}$, when the coupling strength $K$ is slightly larger than $K_c$.  A macroscopic number of oscillators are entrained to the macroscopic oscillation, whose frequency is equal to the average frequency $\omega_0$ of the natural frequencies. The ratio $r$ of the entrained oscillators over the total number $N$ also increases as $r\propto (K-K_c)^{1/2}$. The ratio $r$ is interpreted as an order parameter for the frequency order. 
Strogatz and Mirollo discussed the stability of the disorder phase characterized by $\sigma=0$ for $K<K_c$.~\cite{rf:7}  Crawford derived an amplitude equation for the order parameter $\sigma(t)$ near $k_c$.~\cite{rf:8}  Recently, the Kuramoto model has been intensively studied and an review article of the model was written by Acebr\'on et al.~\cite{rf:9}  

Various generalized coupled phase oscillator models have been studied. 
Sakaguchi and Kuramoto studied a nonvariational model 
\begin{equation}
\frac{d\phi_i}{dt}=\omega_i+\frac{K}{N}\sum_{j=1}^N\{\sin(\phi_j-\phi_i-\alpha)+\sin\alpha\},
\end{equation}
where $\alpha$ is a parameter for the nonvariational coupling.~\cite{rf:10}  
This type of coupling characterized by nonzero $\alpha$ appears naturally in the general phase description of coupled limit-cycle oscillators.~\cite{rf:11} 
Wiesenfeld, Colet, and Strogatz derived this type model equation for coupled Josephson junctions.~\cite{rf:12} 
The coupled Josephson junction model was further studied by Sakaguchi and Watanabe.~\cite{rf:13}
The model with $\alpha=0$ is simpler and analogous to the XY model in magnetic systems, however, the nonvariational model with nonzero $\alpha$ is more natural as a nonequilibrium model.   
This generalized model (2) is also soluble and exhibits a phase transition, however, the frequency of the macroscopic oscillation is generally larger than the average value $\omega_0$ of the natural frequency. It is often observed in realistic systems that the actual frequency, which appears as a result of mutual entrainment, is larger than the average value of the natural frequencies. 
 Daido studied even more generalized coupled phase oscillator models:  
\begin{equation}
\frac{d\phi_i}{dt}=\omega_i+\frac{K}{N}\sum_{j=1}^N h(\phi_j-\phi_i),
\end{equation}
and showed that the order parameter $\sigma$ can increase as $\sigma\sim (K-K_c)$, when $h(\psi)$ includes the second harmonics such as $\sin(2\psi)$.~\cite{rf:14}
We proposed another soluble model including a noise term~\cite{rf:15}:  
\begin{equation}
\frac{d\phi_i}{dt}=\omega_i+\frac{K}{N}\sum_{j=1}^N\sin(\phi_j-\phi_i)+\xi_i(t),\end{equation}
where $\xi(t)$ is a Gaussian white noise satisfying $\langle \xi_i(t)\xi_j(t^{\prime})\rangle=2D\delta_{i,j}\delta(t-t^{\prime})$. The noise $\xi_i(t)$ for the $i$th oscillator is independent of the noise $\xi_j(t)$ for the $j$th oscillator in this model. 
We also studied the synchronization in a population of coupled active rotators: \begin{equation}
\frac{d\phi_i}{dt}=\omega_i-b\sin\phi_i+\frac{K}{N}\sum_{j=1}^N\sin(\phi_j-\phi_i). 
\end{equation}
When the parameter $b$ is increased up to $\omega_i$, each oscillator behaves like a relaxation oscillator, which is characterized by the repetition of long-time slow motion and short-time fast motion. It is because the phase motion becomes slow near $\phi=\pi/2$.  If the parameter $b$ is further increased to be larger than $\omega_i$, the phase locking occurs near $\phi=\pi/2$ and the oscillator changes into an excitable element. The relaxation oscillation and the excitability are also observed in various actual systems. 
In a population of these coupled active rotators, it was shown that the frequency entrainment takes place at several frequencies $\omega=\omega_0+n\Delta \omega_1$. ($n$ is an integer.)~\cite{rf:15}  

We further studied  a slightly modified model equation: 
\begin{equation}
\frac{d\phi_i}{dt}=\omega-b\sin\phi_i+F\sin\omega_0 t+\xi_i(t)+\frac{K}{N}\sum_{j=1}^N\sin(\phi_j-\phi_i),
\end{equation}
in the parameter region $\omega\sim b$.~\cite{rf:16} In this model, external noises and a periodic force are further applied to the coupled active rotator model Eq.~(5). 
When $b\sim \omega$, each element is close to the excitable system. 
Excitable elements under a periodic force and external noises can exhibit the stochastic resonance, that is, the response to the periodic force is enhanced by the external noises of intermediate strength.~\cite{rf:17} In our model, each element is further coupled with the other elements with equal strength.  The stochastic resonance is expected to be enhanced by the cooperative interaction. Indeed, we found that the order parameter exhibits a limit-cycle oscillation with large amplitude at an intermediate strength of external noises, which is a phenomenon analogous to the stochastic resonance.~\cite{rf:16}   
 
Global coupling is assumed in these coupled phase oscillator models, that is, each oscillator interacts with all the other oscillators with the same coupling strength. The global coupling makes a mean-field analysis possible, therefore, the models can become soluble in some cases. In realistic systems, such a uniform and global coupling is not plausible.  Sakaguchi, Shinomoto and Kuramoto studied locally coupled phase oscillators on square lattices and cubic lattices and discussed the possibility of phase transitions in finite dimensional systems.~\cite{rf:18}  
Each oscillator interacts with oscillators on the nearest-neighbor sites as
\begin{equation}
\frac{d\phi_i}{dt}=\omega_i+K\sum_{j\in K_i}\sin(\phi_j-\phi_i),
\end{equation}
where the summation is taken on the nearest-neighbor sites $K_i$ of the $i$th site. We found that the phase order described by the order parameter $\sigma$ does not appear even on the cubic lattices, but the frequency order denoted by the number ratio $r$ of entrained oscillators seems to exist in the three dimensional system. Hong et al. studied numerically the finite dimensional model up to $d=6$ using the finite-size scaling and found the lower critical dimension for the phase order is $d_f=4$.~\cite{rf:19}

Recently, complex networks such as the small-world network~\cite{rf:20} and the scale-free network~\cite{rf:21} have intensively been studied.  In this trend, 
the coupled phase oscillators with variational interaction $\alpha=0$ have been studied on complex networks such as small-world networks and scale-free networks. 
Hong et al. studied the synchronization on the small-world network~\cite{rf:22}, and  Moreno and Pacheco studied the synchronization on the scale-free network, and they found a desynchronization-synchronization transition similar to the one found in the globally coupled system~\cite{rf:23}.  Ichinomiya suggested that the critical coupling strength $K_c$ becomes zero in random scale-free networks with link number distribution $P(l)\propto l^{-\gamma}$ of $2<\gamma\le 3$.~\cite{rf:24} Lee studied the system with the mean-field method.~\cite{rf:25}
Hong et al. studied finite-size scaling near the critical point on complex networks.~\cite{rf:26} 
 
In this paper, we study two new coupled phase oscillator models, which are closely related to the previous models. One model is globally coupled active rotators under common external noises, and the other is coupled phase oscillators with nonvariational interaction of nonzero $\alpha$ on a scale-free network.  

\section{Common-noise-induced synchronization}
Various noise effects such as the stochastic resonance have been intensively studied.~\cite{rf:17} It is known that common noise can induce complete synchronization even for uncoupled homogeneous oscillators.~\cite{rf:27,rf:28} 
Mori and Kai observed noise-induced entrainment in brain waves of humans.~\cite{rf:29} In their experiment, mutual synchronization of $\alpha$ waves was induced by a common noise applied through one eye.   We studied a one-dimensional active-rotator system with unstable coupling under a common noise, and found that mutual synchronization occurs by the common noise. The noise-induced synchronization was most facilitated at an intermediate noise strength, which is analogous to the stochastic resonance.~\cite{rf:30} We found that desynchronization occurs intermittently near a critical parameter where the complete synchronization breaks down. In this section, we study a globally coupled active-rotator system with random natural frequencies under a common noise. The model equation is
\begin{equation}
\frac{d\phi_i}{dt}=\omega_i-b\sin\phi_i+\frac{K}{N}\sum_{j=1}^N\sin(\phi_j-\phi_i)+\xi(t), 
\end{equation}
\begin{figure}[tbp]
\includegraphics[width=9cm]{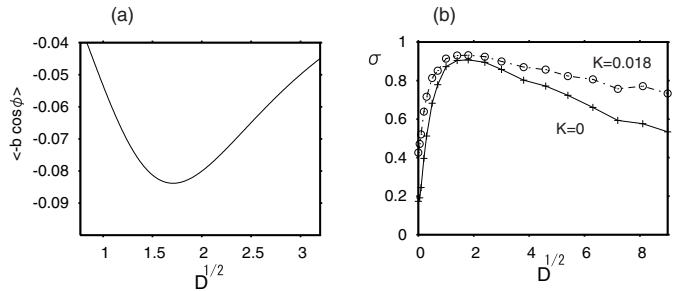}
\caption{(a) Average value of $-b\cos\phi$ at $b=1,\omega_0=3$ as a function of $D^{1/2}$ for the stationary distribution (12). (b) Temporal average of $\sigma$ for model (8) as a function of $D^{1/2}$ at $K=0.018$ (circles) and $K=0$ (crosses) for $N=1000,\omega=3$ and $b=1$.
} \label{fig1}
\end{figure}
\begin{figure}[tbp]
\includegraphics[width=13cm]{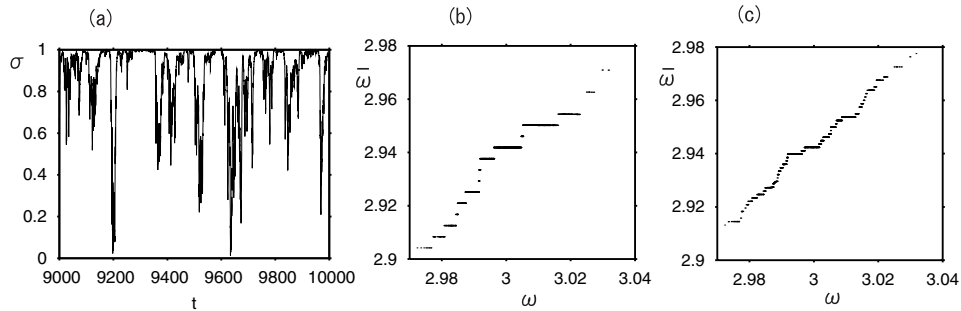}
\caption{(a) Time evolution of $\sigma(t)$ for $K=0,\omega=3,b=1$ and $D^{1/2}=1.8$. (b) Resultant frequency $\bar{\omega}_i=(\phi(t+T)-\phi(t))/T$ vs. $\omega_i$ with $T=1500$. (c) Resultant frequency $\bar{\omega}_i=(\phi(t+T)-\phi(t))/T$ vs. $\omega_i$ with $T=5000$. 
} \label{fig2}
\end{figure}
where $\omega_i$ is a random natural frequency, the probability distribution of $\omega_i$ is assumed to be a Gaussian distribution with the average value $\omega_0$ and the variance $\kappa^2$, and the common noise $\xi(t)$ satisfies $\langle\xi(t)\xi(t^{\prime})\rangle=2D\delta(t-t^{\prime})$.  
This model equation has a form of a mixture of Eq.~(4) and (6), however, the common  noise $\xi(t)$ is applied to all active rotators, and therefore, the subscript $i$ does not appear in the noise $\xi(t)$ in Eq.~(8). If two uncoupled active rotators with the same natural frequency are driven by the common noise, the equations of motion are written as 
\begin{eqnarray}
\frac{d\phi_1}{dt}&=&\omega_0-b\sin\phi_1+\xi(t), \nonumber\\
\frac{d\phi_2}{dt}&=&\omega_0-b\sin\phi_2+\xi(t). 
\end{eqnarray}
In the completely synchronized state, $\phi_1(t)=\phi_2(t)$ is satisfied. The stability of the completely synchronized state is studied by a linearized equation:
\begin{equation}
\frac{d(\delta\phi)}{dt}=-b\cos\phi(t)\delta\phi,
\end{equation} 
where $\delta\phi$ denotes small difference: $\delta \phi=\phi_2(t)-\phi_1(t)$ and $\phi(t)$ denotes the completely synchronized motion: $\phi(t)=\phi_1(t)=\phi_2(t)$. 
The long time average of $-b\cos\phi$ determines the linear stability for the synchronized motion. For $D=0$, the time average of $-b\cos\phi$ is zero, and therefore the synchronized motion is neutrally stable.  For nonzero $D$, it is possible that the long time average of $-b\cos\phi$ becomes negative and the synchronized motion becomes stable.  To calculate the average of $-b\cos\phi$, we use the Fokker-Planck equation corresponding to the Langevin equation Eq.~(9). 
For nonzero $D$, the probability distribution of $\phi$ obeys the Fokker-Planck equation:
\begin{equation}
\frac{\partial P}{\partial t}=-\frac{\partial}{\partial \phi}\{(\omega_0-b\sin\phi)P\}+D\frac{\partial^2P}{\partial \phi^2}.
\end{equation}
The stationary solution $P_s$ is given by \cite{rf:30}
\begin{equation}
P_s(\phi)=P_s(0)\exp\{(\omega_0\phi-b+b\cos\phi)/D\}\left \{1+\frac{(e^{-2\pi\omega_0/D}-1)\int_0^{\phi}e^{(-\omega_0\psi-b\cos\psi)/D}d\psi}{\int_0^{2\pi}e^{(-\omega_0\psi-b\cos\psi)/D}d\psi}\right \},
\end{equation}
where $P_s(0)$ is determined by the normalization condition. 
The long time average of $-b\cos\phi(t)$ can be replaced by the ensemble average by the stationary distribution $P_s(\phi)$. Figure 1 displays the average value of $-b\cos\phi$ as a function of $D^{1/2}$ for $\omega_0=3$ and $b=1$. The average value of $-b\cos\phi$ is indeed negative, and it takes a minimum value at $D^{1/2}\sim 1.7$, where the stability of the synchronized state is the strongest. In other word, the synchronization by the common noise is expected to takes place most strongly near $D^{1/2}=1.7$. 
 
We have performed direct numerical simulation of the coupled phase oscillator model (8) with $N=1000, b=1,\omega_0=3$ and $\kappa=0.01$. Figure 1(b) displays the temporal average of $\sigma=|1/N\sum_{j=1}^N\exp(i\phi_j)|$ as a function of $D^{1/2}$ for $K=0$ and $K=0.018$. The order parameter represents a degree of the synchronization among the phase oscillators. The order parameter takes a maximum value near $D^{1/2}\sim 1.7$, as is expected from the stability exponent. 
These simulations show that the common noises induce the synchronization among active rotators even if the natural frequencies are randomly distributed. The synchronization is most effective at an intermediate strength of noise, which is also analogous to the stochastic resonance.  The value of the order parameter takes a larger value for $K=0.018$ than $K=0$ for any $D$, because the effect of the mutual entrainment is further added in case of nonzero $K$. However, the noise induced synchronization is more clearly seen in the uncoupled system $K=0$, in that the difference of the maximum and the minimum values of the order parameter as a function of $D$ is larger for $K=0$ than $K=0.018$. Figure 2(a) displays a time evolution of $\sigma(t)$ for $K=0$ and $D^{1/2}=1.8$. The time evolution is strongly intermittent. It is because the stability exponent $-b\cos\phi$ fluctuates in time markedly owing to the common noise.  That is, $-b\cos\phi(t)$ is negative on the average and the order parameter tends to take a nonzero value, however, it is possible that $-b\cos\phi(t)$ takes a positive value for a rather long duration, then, the synchronized motion becomes unstable and the order parameter is decreased to nearly zero.     
Figures 2(b) and (c) displays a relation of the numerically obtained resultant frequency $\bar{\omega}_i=(\phi(t+T)-\phi_i(t))/T$ ((b) $T=1500$ and (c) $T=5000$) vs. $\omega_i$ for $K=0$ and $D^{1/2}=1.8$.  Step structures including many fine steps are seen in these figure, which imply mutual frequency entrainment at many different frequencies $\omega_i$ for a finite value of $T$.  Thus, we have shown using this model that the noise-induced synchronization occurs even in a population of active rotators with random natural frequencies. The step structure becomes finer and finer as $T$ is increased as shown in Figs.~2(b) and (c) and the structure disappears at $T\rightarrow \infty$, because the complete frequency entrainment is impossible. However, the phase synchronization characterized by nonzero $\sigma(t)$ is maintained as shown in Fig.~2(a). 
\section{coupled phase oscillators with nonvariational coupling on a scale-free network}
In this section, we study a population of coupled phase oscillators with nonvariational coupling on a scale-free network. The model equation is written as 
\begin{equation}
\frac{d\phi_i}{dt}=\omega_i+K\sum_{j\in K_i}(\sin(\phi_j-\phi_i-\alpha)+\sin\alpha),
\end{equation}
where each oscillator is set on a scale-free network and the summation of the mutual interaction is taken only between oscillators connected by links in the scale-free network.  We studied  such nonvariational systems on linear, square and cubic lattices, and found that phase waves such as target waves and spiral waves appear inside of mutually entrained domains in the finite-dimensional nonvariational systems.~\cite{rf:31} The nonvariational coupling is essential for the formation of the phase waves. The phase waves play very important roles for mutual entrainment. As a result, the mutual entrainment in systems with nonzero $\alpha$ can be rather different from that in the system of $\alpha=0$. 
For example, we found the multi-stability of mutual entrainment in a system of $\alpha=\pi/4$ on a square lattice.~\cite{rf:31} 

Moreno and Pacheco~\cite{rf:23} studied coupled phase oscillators with variational interaction of $\alpha=0$ on a scale-free network proposed by Barab\'asi and Arbert.~\cite{rf:21} The scale-free network by Barab\'asi and Arbert has a power-law degree distribution $P(l)\sim l^{-\gamma}$ with $\gamma=3$. Moreno and Pacheco found that the order parameter increases continuously from zero as $K$ is increased. They analyzed the system by a finite-size scaling method and found that a phase transition occurs at a finite $K$. The order parameter increases $\sigma\sim (K-K_c)^{\beta}$ with $\beta=0.46$, which is consistent with the critical exponent 0.5 in the original globally coupled system (1). 
However, Lee suggested that the exponent $\beta$ changes as $\beta=1/(\gamma-3)$, which becomes $\infty$ at $\gamma=3$. We have performed similar numerical simulations of coupled phase oscillators with nonvariational interaction for $N=2000$. The preferential attachment algorithm by Barab\'asi and Arbert is used to construct a scale-free network. There is a free parameter $m$ in the Barab\'asi-Albert model. The parameter $m$ denotes the number of older sites to which a newly generated node attaches. In a scale-free network with parameter $m$, the average link number is expressed as $\langle l\rangle\sim 2m$. 
We have first performed numerical simulations for a scale-free network with $m=2$, and the probability distribution of $\omega$ is assumed to be a Gaussian distribution of $\kappa=0.1$ and $\omega_0=0$.

\begin{figure}[tbp]
\includegraphics[width=15cm]{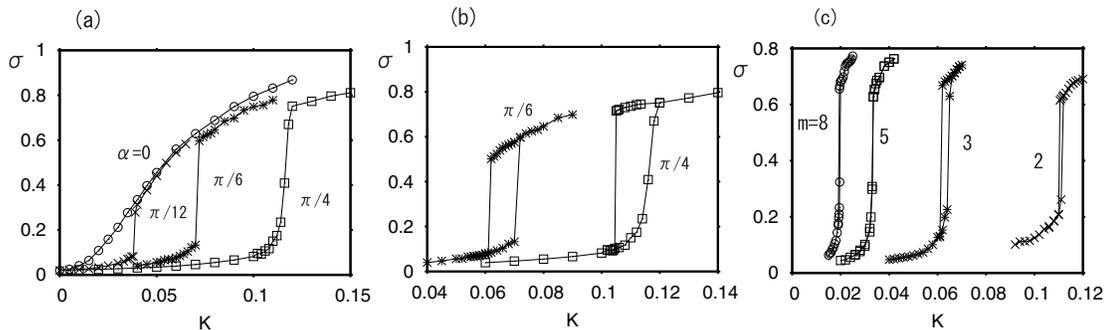}
\caption{(a) Order parameter $\sigma$ as a function of $K$ for $\alpha=0,\pi/12,\pi/6$ and $\pi/4$. (b) Hysteresis of the order parameter for $\alpha=\pi/6$ and $\pi/4$. (c) Order parameter $\sigma$ as a function of $K$ for $\alpha=\pi/4$ on the scale-free networks of $m=2,3,5$ and $8$.
} \label{fig3}
\end{figure}
Figure 3(a) displays $\sigma$ as a function of $K$ for $\alpha=0,\pi/12,\pi/6$ and $\pi/4$. The coupling constant $K$ is stepwise increased in the numerical simulation, that is, a final state of ${\phi_i}$ at a coupling constant $K$ is used as a fresh initial condition at a slightly larger coupling constant $K+\Delta K$. For $\alpha=0$, we found a continuous transition which is similar to that investigated by Moreno and Pacheco. However, the phase transition becomes sharp at $\alpha=\pi/12,\pi/6$ and $\alpha=\pi/4$. The critical values are evaluated respectively at $K_c=0.039, 0.071$ and $0.114$ for $\alpha=\pi/12,\pi/6$ and $\pi/4$. 
Figure 3(b) displays $\sigma$ as a function of $K$ for $\alpha=\pi/6$ and $\pi/4$, however, in this simulation, the coupling constant $K$ is stepwise increased, and then $K$ is inversely stepwise decreased.  For these parameter values of $\alpha$, the transitions are discontinuous and a hysteresis appears, although the bistable range is rather small. The bistability was not clearly observed for $\alpha=\pi/12$. We have checked that the sharp transitions occur also for other scale-free networks with $m=2,3,5,$ and 8.  Figure 3(c) displays the order parameter $\sigma$ when the coupling strength $K$ is changed stepwise at $\alpha=\pi/4$. (Here, a different scale-free network from the one used in Figs.3(a) and (b) was constructed for $m=2$.) We can evaluate the critical coupling strength $K_c$ as 0.1105,0.0625,0.033 and 0.0195 respectively for $m=2,3,5$ and 8. As $m$ is increased, $\langle l\rangle K_c=2mK_c$ seems to approach a constant value about 0.29.

In the scale-free network, there are several hub sites which might play an important role 
even for the mutual entrainment. In the coupled oscillators with nonvariational interaction $\alpha>0$, oscillators with larger connections tend to have larger frequency, if mutual synchronization does not occur. It is because the interaction term of the right-hand side of Eq.~(13) is decomposed into
\begin{equation}
K\sum_{j\in K_i}\{\sin(\phi_j-\phi_i-\alpha)+\sin\alpha\}=K\sum_{j\in K_i}\cos\alpha\sin(\phi_j-\phi_i)+K\sum_{j\in K_i}\sin\alpha\{1-\cos(\phi_j-\phi_i).\}
\end{equation}
The second term of the right-hand side of Eq.~(14) makes the resultant frequency increase.  Because oscillators at hub sites have a large number of connections and the resultant frequencies@increase owing to the summation term, which makes the mutual entrainment difficult at small $K$ for the hub oscillators. 
\begin{figure}[tbp]
\includegraphics[height=3.5cm]{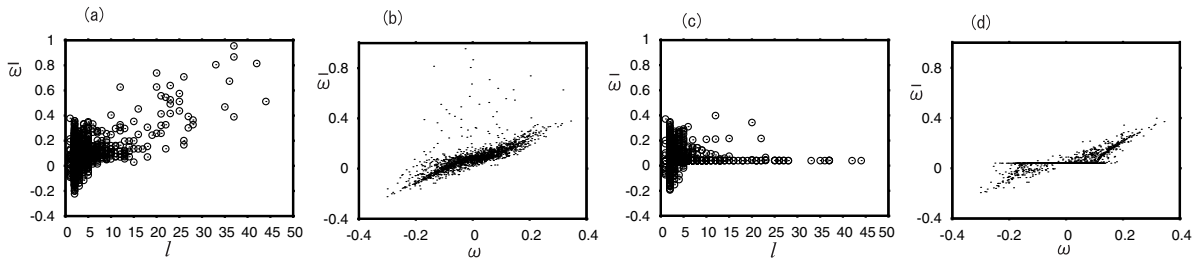}
\caption{(a) Resultant frequency $\bar{\omega}$ vs. link number $l_i$ at $m=2,\alpha=\pi/6$ and $K=0.07$ in a desynchronized state. (b) Resultant frequency $\bar{\omega}$ vs. natural frequency $\omega$ in the desynchronized state. (c) Resultant frequency $\bar{\omega}$ vs. link number $l_i$ in a synchronized state. (d) Resultant frequency $\bar{\omega}$ vs. natural frequency $\omega$ in the synchronized state.}
\label{fig4}
\end{figure}
\begin{figure}[tbp]
\includegraphics[height=4.cm]{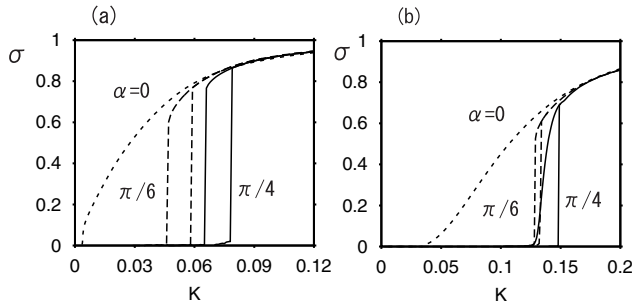}
\caption{(a) Order parameter $\sigma$ as a function of $K$ numerically obtained  from the self-consistent equation (16) at $\alpha=0$ (dotted curve), $\alpha=\pi/6$ (dashed curve) and $\alpha=\pi/4$ (solid curve) for $\kappa=0.1$ and $P_1(l)$. (b) Order parameter $\sigma$ as a function of $K$ at $\alpha=0$ (dotted curve), $\alpha=\pi/6$ (dashed curve) and $\alpha=\pi/4$ (solid curve) for $\kappa=0.1$ and $P_2(l)$.}
\label{fig5}
\end{figure}
However, for sufficiently large $K$, mutual synchronization proceeds and the phase differences $|\phi_j-\phi_i|$ become small, and the effect by the term $\sum \sin\alpha\{1-\cos(\phi_j-\phi_i)\}$ becomes weak, then  the mutual synchronization rapidly extends via the hub oscillators. It might be a reason of the bistability and the hysteresis.  Figures 4(a) displays a relation of the link number $l_i$ and the resultant frequency $\bar{\omega}_i$ for each oscillator, and Figure 4(b) a relation of the natural frequency $\omega_i$ and the resultant frequency $\bar{\omega}_i$ at $\alpha=\pi/6$ and $K=0.07$ in a desynchronized state with $\sigma\sim 0.13$. The scale-free network of $m=2$ is used. It is clearly seen in Fig.~4(a) that the resultant frequencies are large for oscillators with large link number. No clear frequency locking is seen in Fig.~4(b).   Figure 4(c) and (d) display a relation of $l_i$ and $\bar{\omega}_i$,  and  a relation of  $\omega_i$ and $\bar{\omega}_i$ at the same parameter $\alpha=\pi/6$ and $K=0.07$ in a synchronized state with $\sigma\sim 0.58$. In this synchronized state, the oscillators with large link number are well entrained to the frequency $\bar{\omega}\sim 0.0408$. The resultant frequency $\bar{\omega}\sim 0.0408$ is definitely larger than the average value $\omega_0=0$ of the natural frequency. It is characteristic of the oscillator system with nonvariational interaction.  There is a tendency that the phase $\phi_i$  for the oscillators with larger link number $l$ is larger in a population of the mutually entrained oscillators. That is, the hub oscillators tend to be in leading positions of the oscillation or play a role of pacemakers among the mutually-entrained oscillators. This is analogous to the fact that the oscillators with large natural frequencies play a role of pacemakers,  which appear as centers of target waves, in finite-dimensional systems with nonvariational interaction.~\cite{rf:31} 

The mean-field  analysis might be applicable also in this system, if the phase correlation owing the network structure is neglected.~\cite{rf:10,rf:24,rf:25,rf:26} The order parameter weighed with the link number is defined as 
\[\sigma e^{i\alpha}=\sum_{l=1}^{\infty}P(l)le^{i\psi(l)}/\sum_{l=1}^{\infty} P(l)l,\]
where $P(l)$ is the probability distribution of the link number $l$, another phase $\psi$ is introduced as $\psi=\phi-\Omega t+\alpha$, and $\Omega$ is the frequency of the order parameter. The phase $\psi_i(l)$ of the $i$th oscillator with link number $l$  obeys approximately the equation:
\begin{equation}
\frac{d\psi_i}{dt}=\omega_i-\Omega+Kl\sin\alpha-Kl\sigma\sin\psi_i.
\end{equation}
If $|\omega_i-\omega+Kl\sin\alpha|$ is smaller than $Kl$, the oscillator is entrained to the mean-field and the stationary solution of Eq.~(15) is $\psi_i(l)=\sin^{-1}\{(\omega_i-\omega+Kl\sin\alpha)/Kl\}$. Since the natural frequency obeys the probability distribution $g(\omega)$, the phase distribution $n_s(\psi;l)$ of the entrained oscillator is expressed as 
\[n_s(\psi;l)=g(\Omega-Kl\sin\alpha+Kl\sigma\sin\psi)Kl\sigma\cos\psi,\;\;|\psi|\le \pi/2.\]
If $|\omega_i-\omega+Kl\sin\alpha|$ is larger than $Kl$, the oscillator is not entrained to the mean-field, then the phase distribution $n_{ds}(\psi;\omega,l)$  of the desynchronized oscillators is expressed as 
\[n_{ds}(\psi;\omega,l)\propto 1/|\omega-\Omega+Kl\sin\alpha-Kl\sigma\sin\psi|.\]
The order parameter can be expressed using the two phase distributions $n_s$ and $n_{ds}$. Then, we get a self-consistent equation for the order parameter and the frequency $\Omega$ as 
\begin{equation}
\sigma e^{i\alpha}\sum P(l) l=\sum P(l)Kl^2\sigma\left\{\int_{\pi/2}^{\pi/2}d\psi g(\Omega-Kl\sin\alpha+Kl\sigma\sin\psi)\cos\psi e^{i\psi}+iJ(l)\right\},
\end{equation}  
where
\[J(l)=\int_0^{\pi/2}d\psi\frac{\cos\psi(1-\cos\psi)}{\sin^3\psi}\left\{g(\Omega-Kl\sin\alpha+Kl\sigma/\sin\psi)-g(\Omega-Kl\sin\alpha-Kl\sigma/\sin\psi)\right\}.\] 
If $\alpha=0$, $\Omega=0$ and the critical value $K_c$ is calculated as $K_c=2\langle l\rangle/\{\pi g(0)\langle l^2\rangle\}$. If $P(l)\propto 1/l^3$ as in our scale-free network, $K_c$ become zero because $\langle l^2\rangle \propto \sum_{l=1}^{\infty} (l^2/l^3)=\infty$. However, if the system size is finite, the maximum number $l_m$ of the link number is finite. In that case, the critical value $K_c$ is not zero, and it is not so small because $\langle l^2\rangle$ diverges weakly as $\ln l_m$.  
If $\alpha$ is not zero and $\sigma$ is assumed to be infinitesimal, Eq.~(16) is reduced to 
\begin{eqnarray}
\cos\alpha\langle l\rangle&=&K\sum P(l) l^2 (\pi/2)g(\Omega-Kl\sin\alpha),\nonumber\\
\sin\alpha\langle l\rangle&=&K\sum P(l) l^2J_1(\Omega-Kl\sin\alpha),
\end{eqnarray}
where $J_1(\Omega)=\int_0^{\infty}dx\{g(\Omega+x)-g(\Omega-x)\}/(2x)$.~\cite{rf:10} The critical values $K_c$ and $\Omega_c$ are obtained as a solution of the coupled equations (17). 

We have numerically solved the self-consistent equation (16) using two link number distributions, i.e., $P_1(l)$ calculated from the numerically constructed scale-free network for $N=2000$ and $m=2$ used in the previous numerical simulations shown in Fig.~3 and 4, and $P_2(l)\propto 1/l^3$ with $l_m=1000$. The natural frequency distribution $g(\omega)$ is assumed to be the Gaussian distribution  with $\kappa=0.1$ and $\omega_0=0$.  The average link number $\langle l\rangle$ is 3.94 for $P_1(l)$ and $1.37$ for $P_2(l)$.
The self-consistent equation was solved by an iteration method. Figures 5(a) and (b) display $\sigma$ as a function of $K$ at $\alpha=0$ (dotted curve), $\pi/6$ (dashed  curve) and $\alpha=\pi/4.$ (solid curve) for (a) $P_1(l)$ and (b) $P_2(l)$. For $\alpha=0$, continuous transitions from the desynchronized state to the synchronized state occur at $K_c=0.0037$ for $P_1(l)$ and $K_c=0.035$ for $P_2(l)$. The critical values and the order parameter $\sigma$ sensitively depend on the link number distribution. Here, the distribution $P_2(l)$ decreases monotonously as $1/l^3$.  However, in the numerically constructed network characterized by $P_1(l)$, the largest hub site has a link number $l_i=430$ and the second-largest hub site  has a link number $l_i=249$, although $P_1$ is roughly proportional $1/l^3$ for $l<20$. For $P_2(l)$, the exponent $\beta$ in $\sigma\propto (K-K_c)^{\beta}$ is rather large as was suggested by Lee,~\cite{rf:25}, but $\beta$ is about 0.5 for $P_1(l)$, which is consistent with  our direct numerical simulation and the numerical simulation by Moreno and Pacheco. For $\alpha=\pi/6$, discontinuous transitions from the synchronized state to the desynchronized state occur at $K=0.047$ for $P_1(l)$ and at $K=0.128$ for $P_2(l)$, when $K$ is continuously decreased.  For $\alpha=\pi/4$, discontinuous transitions from the synchronized state to the desynchronized state occur at $K=0.065$ for $P_1(l)$ and at $K=0.128$ for $P_2(l)$, when $K$ is continuously decreased.   Transitions from the desynchronized state to the synchronized state occur respectively at $K=0.059$ and $K=0.079$ for $\alpha=\pi/6$ and $\pi/4$ in Fig.~5(a),  and at $K=0.134$ and $K=0.149$ for $\alpha=\pi/6$ and $\pi/4$ in Fig.~5(b), when $K$ is continuously increased. That is, hysteresis is seen also in the mean-field analysis. For $\alpha=\pi/4$ and $P_1(l)$, the average of the two critical values is estimated as $K_c\sim 0.072$, then, $\langle l\rangle K_c\sim 0.285$, which is close to the critical value for large $m$ in Fig.~3(c). 
These results of the mean-field approximation for $P_1(l)$ are qualitatively consistent with the results of the direct numerical simulation shown in Fig.~3.  However, quantitative agreement is not so satisfactory, probably because the mean-field approximation is not exact on the scale-free network and the finite-size effect is rather strong. 
\section{Summary}
 In this report, we have made a short review about the Kuramoto model and the generalization. Next, we have studied the common-noise-induced synchronization among active rotators. At an intermediate noise strength, the noise-induced synchronization takes place most effectively, which is analogous to the stochastic resonance.   Finally, we have studied the synchronization of coupled phase oscillators with nonvariational interaction on scale-free networks. We have found sharp transitions and weak hysteresis in the nonvariational systems. The mutual entrainment via hub oscillators seems to be important in these nonvariational systems.
We have numerically solved the self-consistent equation by the mean-field approximation. We have found that the discontinuous transition occurs also in the mean-field approximation for nonzero $\alpha$.  


\begin{references}
\bibitem{rf:1} A.~Pikovsky, M.~Rosenblum and J.~Kurths, {\it Synchronization: A universal Concept in Nonlinear Sciences} (Cambridge University Press, Cambridge, 2001).   
\bibitem{rf:2} J.~Buck and E.~Buck, Science, {\bf 159},1319 (1968).
\bibitem{rf:3} S.~Yamaguchi, H.~Isejima, T.~Matsuo, R.~Okura, K.~Yagita, M.~Kobayashi, H.~Okamura, Science {\bf 302}, 1408 (2003).
\bibitem{rf:4}G.~Buzs\'aki {\it Rhythms of the Brain} (Oxford University Press, New York, 2006).
\bibitem{rf:5} A.~T.~Winfree, J. Theor. Biol. {\bf 16}, 15 (1967). 
\bibitem{rf:6} Y.~Kuramoto, in H.~Araki ed. {\it Lecture Notes in Physics} (Springer-Verlag, New-York) {\bf 39}, 420 (1975). 
\bibitem{rf:7} S.~H.~Strogatz and R.~Mirollo, J. Stat. Phys. {\bf 63}, 613 (1991).
\bibitem{rf:8} J.~D.~Crawford, Phys. Rev. Lett. {\bf 74}, 4341 (1995).
\bibitem{rf:9} J.~A.~Acebr\'on, L.~L.~Bonilla, C.~J.~P\'erez Vicente, F.~Rotort and R.~Spigler, Rev. Mod. Phys. {\bf 77}, 137 (2005).
\bibitem{rf:10} H.~Sakaguchi and Y.~Kuramoto, Prog. Theor. Phys. {\bf 76},576 (1986).
\bibitem{rf:11} Y.~Kuramoto, {\it Chemical Oscillations, Waves, and Turbulence} (springer-Verlag, New York, 1984).
\bibitem{rf:12} K.~Wiesenfeld, P.Colet and S.~H.~Strogatz, Phys. Rev. E {\bf 57}, 1563 (1998).
\bibitem{rf:13} H.~Sakaguchi and K.~Watanabe, J. Phys. Soc. Jpn. {\bf 69}, 3545 (2000).
\bibitem{rf:14} H.~Daido, Phys. Rev. Letts. {\bf 73}, 760 (1994).
\bibitem{rf:15} H.~Sakaguchi, Prog. Theor. Phys. {\bf 79}, 39 (1988).
\bibitem{rf:16} H.~Sakaguchi, Phys. Rev. E {\bf 66}, 056129 (2002).
\bibitem{rf:17} L.~Gammaitoni, P.~H\'anggi, P.~Jung and F.~Marchensoni, Rev. Mod. Phys. {\bf 70}, 223 (1998). 
\bibitem{rf:18} H.~Sakaguchi, S.~Shinomoto and Y.~Kuramoto, Prog. Theor. Phys. {\bf 77}, 1005 (1987).
\bibitem{rf:19} H.~Hong, H.~Park and M.~Y.~Choi, Phys. Rev. E {\bf 70}, 045204(R) (2004). H.~Hong, H.~Chat\'e, H.~Park and L-H.~Tang, Phys. Rev. Lett. {\bf 99}, 184101 (2007).
\bibitem{rf:20} D.~J.~Watts and S.~H.~Strogatz, Nature {\bf 393}, 440 (1998).
\bibitem{rf:21} A.~L.~Barab\'asi and R.~Albert, Science {\bf 286}, 509 (1999).
\bibitem{rf:22} H.~Hong, M.~Y.~Choi and B.~J.~Kim, Phys. Rev. E {\bf 65}, 026139 (2002).
\bibitem{rf:23} Y.~Moreno and A.~F.~Pacheco, Europhys. Lett. {\bf 68}, 603 (2004).
\bibitem{rf:24} T.~Ichinomiya, Phys. Rev. E {\bf 70}, 026116 (2004).
\bibitem{rf:25} D-S.~Lee, Phys Rev. E {\bf 72}, 026208 (2005).
\bibitem{rf:26} H.~Hong, H.~Park and L-H.~Tang, Phys. Rev. E {\bf 76}, 066104 (2007).
\bibitem{rf:27} A.~Maritan and J.~R.~Banavar, Phys. Rev. Lett. {\bf 72}, 1451 (1994).
\bibitem{rf:28} J.~Teramae and D.~Tanaka, Phys. Rev. Lett. {\bf 93}, 204103 (2004).
\bibitem{rf:29} T.~Mori and S.~Kai, Phys. Rev. Lett. {\bf 88}, 218101 (2002).
\bibitem{rf:30} H.~Sakaguchi, Phys. Letts. A {\bf 318}, 553 (2003).
\bibitem{rf:31} H.~Sakaguchi, S.~Shinomoto and Y.~Kuramoto, Prog. Theor. Phys. {\bf 79}, 1069 (1989).
\end{references}
\end{document}